\documentstyle[12pt]{article}

\newlength{\dinwidth}
\newlength{\dinmargin}
\def\lapproxeq{\lower .7ex\hbox{$\;\stackrel{\textstyle
<}{\sim}\;$}}
\def\gapproxeq{\lower .7ex\hbox{$\;\stackrel{\textstyle
>}{\sim}\;$}}

\begin{document}
\titlepage
\begin{flushright}
DTP/95/94  \\
October 1995 \\
hep-ph/9511263 \\
\end{flushright}

\begin{center}
\vspace*{2cm}
{\Large \bf The description of $F_2$ at small $x$ incorporating
angular ordering} \\
\vspace*{1cm}
J.\ Kwieci\'{n}ski\footnote{On leave from Henryk
Niewodnicza\'{n}ski Institute of Nuclear Physics, 31-342
Krak\'{o}w, Poland.}, A.\ D.\ Martin and P.\ J.\ Sutton,

Department of Physics, University of Durham, Durham, DH1 3LE,
England
\end{center}

\vspace*{5cm}
\begin{abstract}
We study the perturbative QCD description of the HERA
measurements of $F_2 (x, Q^2)$ using a gluon distribution that is
obtained from an evolution incorporating angular ordering of the
gluon emissions, and which embodies both GLAP and BFKL dynamics.
We compare the predictions with recent HERA data for $F_2$.  We
present estimates of the charm component $F_2^c (x, Q^2)$ and of
$F_L (x, Q^2)$.
\end{abstract}

\newpage

Deep inelastic electron-proton scattering experiments at HERA
have measured the structure function $F_2 (x, Q^2)$ in the
previously unexplored small $x$ regime, $x \lapproxeq 10^{-3}$.
The values of $F_2$ are found to rise rapidly with decreasing $x$
\cite{h1,zeus}.  These measurements have stimulated much
theoretical activity and the small $x$ behaviour of $F_2$ has
been interpreted using perturbative QCD from several different
viewpoints.  The interpretation is complicated by the need to
provide non-perturbative input.

In fact the present data for $F_2$ can be well described by
traditional Altarelli-Parisi (or GLAP) evolution in the
next-to-leading order approximation.  The data imply a steep
gluon (that is a gluon density which increases as $x$ decreases)
even at low $Q^2$ values.  We may input the steep $x$ behaviour
directly into the starting distributions at some input scale, say
$Q_0^2 = 4 \: {\rm GeV}^2$ \cite{mrs}, or alternatively we may
generate it from \lq\lq non-singular" or \lq\lq flat" $x$
distributions at some low scale, such as $Q_0^2 = 0.3 \: {\rm
GeV}^2$ \cite{grv} or $Q_0^2 = 1 \: {\rm GeV}^2$ \cite{bf},
chosen so that the evolution length
$$
\xi (Q_0^2, Q^2) = \int_{Q_0^2}^{Q^2} \: \overline{\alpha}_S
(q^2) \: \frac{dq^2}{q^2}
$$
\noindent is sufficiently long, where $\overline{\alpha}_S \equiv
3 \alpha_S/\pi$.  GLAP evolution amounts to the resummation of
the leading (and next-to-leading) $\log Q^2$ terms.  At small $x$
and large $Q^2/Q_0^2$ it generates a steep double leading
logarithmic (DLL) behaviour\footnote{This form increases with
decreasing $x$ faster than any power of $\log 1/x$ but slower
than any power of $1/x$.  A choice of singular starting
distributions, $xg, xq_{\rm sea} \sim x^{- \lambda}$ with
$\lambda > 0$, would therefore eventually override the DLL
behaviour.} of the form $\exp (2 [ \xi (Q_0^2, Q^2) \log (1/x)
]^{\frac{1}{2}} )$.  Despite the apparent success of the GLAP
approach, it is not the only way of generating a steep gluon
compatible with the HERA data.

At sufficiently small $x$ we must also resum the $\alpha_S \log
1/x$ terms, unaccompanied by $\log Q^2$.  This is accomplished by
the BFKL equation.  It generates a singular $x^{- \lambda_L}$
behaviour of the unintegrated gluon distribution, $f (x, k_T^2)$,
where $\lambda_L = \overline{\alpha}_S 4 \ln 2$ for fixed
$\alpha_S$.  If a reasonable assumption is made to introduce the
running of $\alpha_S$ then the numerical solution of the BFKL
equation again yields an $x^{- \lambda}$ behaviour but with
$\lambda \simeq 0.5$ \cite{akms}.  Using the $k_T$-factorization
theorem the behaviour $f(x, k_T^2) \sim x^{- \lambda}$ feeds
through into $F_2$ (and into $F_L$).  To be precise we have
\begin{equation}
F_i (x, Q^2) = \int \: \frac{d k_T^2}{k_T^2} \: \int_x^1 \:
\frac{dx^\prime}{x^\prime} \: f(x^\prime, k_T^2) \: F_i^{\rm box}
\left (\frac{x}{x^\prime}, k_T^2, Q^2 \right ) + F_i^S
\label{a1}
\end{equation}
\noindent with $i = 2, L$, where $F_i^S \simeq F_i (x, Q^2)$ at
large $x$, but is a slowly varying function of $x$ and $Q^2$ at
small $x$.  The convolution in (\ref{a1}) is diagrammatically
displayed in Fig.\ 1.  $F_i^{\rm box}$ includes both the quark
box and crossed box contributions which originate from virtual
photon-virtual gluon $q\overline{q}$ production.  For the
$c\overline{c}$ component we take the quark mass to be $m_c = 1.4
\: ({\rm or} \: 1.7) \: {\rm GeV}$.  At sufficiently small $x$
the
$x^{- \lambda}$ BFKL behaviour overrides the DLL form.  To find
precisely where this will happen requires a unified BFKL/GLAP
formalism, as well as knowledge of the yet unknown
next-to-leading $\log 1/x$ contributions.

One approach which unifies the GLAP and BFKL formalisms is to
recast the leading twist part of the BFKL $k_T$-factorization
formula into collinear form in which the splitting and
coefficient functions acquire the higher order $\log 1/x$
contributions \cite{ch}.  Several interesting phenomenological
studies have developed from this formalism \cite{ekl,bf2,frt}.
Actually the procedure is to use the BFKL equation for fixed
$\alpha_S$ to obtain $\alpha_S/\omega$ power series expansions of
the anomalous dimensions and coefficient functions, and then to
let $\alpha_S$ run.  Here $\omega$ is the moment index.  In this
way we retain the simplicity of GLAP evolution, but with
splitting and coefficient functions which incorporate BFKL
resummations\footnote{There is freedom in assigning the BFKL
resummations to the coefficient functions, splitting functions
and the starting distributions.  Various factorization schemes
have been proposed \cite{c,cat,bf3}.}.

We can see that the basic quantity is the unintegrated gluon
distribution $f$, which corresponds to the sum of (effective)
gluon ladder diagrams.  The unintegrated gluon satisfies the BFKL
equation
\begin{equation}
f(x, k_T^2) = f^0 (x, k_T^2) + \overline{\alpha}_S \: k_T^2 \:
\int_x^1 \: \frac{dz}{z} \: \int \: \frac{dk_T^{\prime
2}}{k_T^{\prime 2}} \: \left [ \frac{f (z, k_T^{\prime 2}) - f
(z, k_T^2)}{| k_T^{\prime 2} - k_T^2 |} + \frac{f (z, k_T^2)}{(4
k_T^{\prime 4} + k_T^4)^{\frac{1}{2}}} \right ],
\label{a2}
\end{equation}
\noindent where $\overline{\alpha}_S \equiv 3 \alpha_S/\pi$.  We
notice the potential collinear singularity at $k_T^{\prime 2} =
0$.  However, provided the driving term $f^0$ is chosen to vanish
at $k_T^2 = 0$, the structure of the equation guarantees that it
is free of collinear singularities.  This is a natural
way\footnote{In fact the vanishing of the inhomogeneous term is
ensured by the colour neutrality of the probed hadron.} to
regulate the singularity.  It can be linked directly with the
$Q_0$ scheme advocated by Ciafaloni \cite{c}.  Since we stay in
four dimensions we avoid factors which are characteristic of
minimal subtraction (MS) schemes.

Here we work with the $k_T$-factorization formula, (\ref{a1}),
and do not reduce the equation to collinear form.  This allows us
to study the effect of replacing the BFKL gluon with the gluon
obtained from the CCFM equation \cite{ccfm}; a unified equation
which embodies both the BFKL equation at small $x$ and GLAP
evolution at large $x$.  The CCFM equation is based on the
coherent radiation of gluons, which leads to an angular ordering
of the emitted gluons.

Since the next-to-leading $\log 1/x$ contributions are not yet
known, the introduction of running $\alpha_S$ into the BFKL
equation\footnote{We thank M.\ Ciafaloni and L.\ Lipatov for
valuable discussions on this point.} is, of necessity, subject to
assumption.  The most reasonable procedure is to take $\alpha_S
(k_T^2)$ in (\ref{a2}), so that the equation is compatible with
the double-leading-logarithm limit of GLAP evolution.  This
prescription, however, generates a solution which differs from
that obtained when the (fixed $\alpha_S$) BFKL equation is first
reduced to collinear form and then $\alpha_S$ is allowed to run
\cite{km}.  In other words the introduction of $\alpha_S (k_T^2)$
in (\ref{a2}) gives a different solution to that obtained by
allowing $\alpha_S$ to run in the leading-twist collinear
solution of the BFKL equation.  Formally, however, the difference
between the two approaches can be attributed to non-leading $\ln
1/x$ effects.  For instance in the region
$$
\frac{\overline{\alpha}_S (Q^2)}{\omega} \; < \;
\frac{\overline{\alpha}_S (Q_0^2)}{\omega} \; < \; \frac{1}{4 \ln
2}
$$
\noindent both methods give the same result up to these
non-leading effects.  However, the singularity structure in the
moment or $\omega$ plane is different in the two formulations.
The collinear reduction of the BFKL equation, leading to
conventional evolution from, say, $Q_0^2$ to $Q^2$, contains a
branch point singularity at $\omega = \omega_L (Q_0^2) \equiv
\overline{\alpha}_S (Q_0^2) 4 \ln 2$ or, if this leading
singularity is absorbed in the starting distributions, at $\omega
= \omega_L (Q^2)$.  On the other hand the solution with running
$\alpha_S$ directly incorporated into the BFKL equation does not
contain the branch point singularity but rather it has (an
infinite number of) poles in the $\omega$ plane.  The leading
pole is well separated from the others with a position $\omega_p
\approx a \alpha_S (b k_0^2)$ where $a$ and $b$ are constants ($b
\approx 7$) and $k_0$ delimits the infrared region.  It turns
out that $\omega_p < \omega_L (k_0^2)$ \cite{mksl}.

Following ref. \cite{ccfm}, we now implement angular ordering of
the gluon emissions.  The unintegrated gluon distribution is then
a solution of the CCFM equation\footnote{The CCFM equation
incorporates part of the non-leading $\log 1/x$ contributions.}
\cite{ccfm} rather than the BFKL equation.  In the small $x$
region the CCFM equation may be approximated by
\begin{equation}
f (x, k_T^2, \overline{Q}^2) = f^0 (x, k_T^2, \overline{Q}^2) +
\overline{\alpha}_S \: k_T^2 \: \int_x^1 \: \frac{dz}{z} \: \int
\: \frac{d^2 q}{\pi q^2} \Theta (\overline{Q} - zq) \: \Delta_R
(z, q, k_T) \: \frac{1}{k_T^{\prime 2}} \: f \left (
\frac{x}{z}, k_T^{\prime 2}, q^2 \right )
\label{a3}
\end{equation}
\noindent with $\mbox{\boldmath $k$}_T^\prime = \mbox{\boldmath
$k$}_T + \mbox{\boldmath $q$}$, see eq.\ (18) of ref.\
\cite{kms}.  $\Delta_R$ represents the virtual corrections which
screen the $1/z$ singularity and the theta function imposes the
angular ordering on the real emissions.  Eq.\ (11) of ref.\
\cite{kms} gives the explicit expression for $\Delta_R$.
We note that the solution $f$ depends on an additional scale
$\overline{Q}$ that is required to specify the maximum angle of
gluon emission (which turns out to be essentially the scale
$\kappa$ of the probe, see Fig.\ 1).  This equation has recently
been solved numerically and the resulting gluon distribution has
been compared with that obtained from the BFKL equation
\cite{kms}.  As anticipated, the angular ordering constraint
suppresses the CCFM gluon at the lower $\overline{Q}^2 =
\kappa^2$ values.  If we replace the angular-ordering constraint
$\Theta (\overline{Q} - zq)$ by $\Theta (\overline{Q} - q)$ and
set $\Delta_R = 1$ then we obtain an equation which becomes
equivalent to the Altarelli-Parisi (GLAP) equation in the
double-leading-logarithm approximation (DLLA).

The small $x$ approximation of the CCFM equation that we have
used (see \cite{kms}) amounts to setting the Sudakov form factor
$\Delta_S = 1$ and to approximating the gluon-gluon splitting
function by its singular term as $z \rightarrow 0$, that is
$P_{gg} \simeq 6/z$.  $\Delta_S$ represents the virtual
corrections which cancel the singularities at $z = 1$.  We
account for the remaining finite terms in $P_{gg}$ by multiplying
the solution $f (x, k_T^2, \overline{Q}^2)$ by the factor
\begin{equation}
\exp \left ( - A \: \int^{\overline{Q}^2} \: \overline{\alpha}_S
(q^2) \: \frac{dq^2}{q^2} \right )
\label{a4}
\end{equation}
\noindent where $A$ is defined by
\begin{equation}
\int_0^1 \: z^\omega \: P_{gg} (z) dz \simeq \frac{6}{\omega} -
6A.
\label{a5}
\end{equation}
\noindent That is $A = (33 + 2n_f)/36$, where the number of
active flavours $n_f = 4$.

Fig.\ 2 compares the CCFM and DLLA predictions for $F_2$ with the
recent HERA measurements \cite{h1,zeus}.  The predictions are
obtained by first determining the gluon distribution $f (x,
k_T^2, \overline{Q}^2)$ by iteration of (\ref{a3})\footnote{The
DLLA prediction is obtained as defined above, that is by taking
$\Delta_R = 1$, and $z = 1$ in the $\Theta$ function in
(\ref{a3}).} in the domain $k_T^2 > k_0^2 = 1 \: {\rm GeV}^2$
starting from a \lq\lq flat" driving term of the form $3 (1 -
x)^5 \exp (-k_T^2/k_a^2)$ with $k_a^2 = 1 \: {\rm GeV}^2$, that
is exactly as in ref.\ \cite{kms}.  We correct for the small $x$
approximation by multiplying the gluon distribution $f$ by the
factor shown in (\ref{a4}) and then predict $F_2$ from the
$k_T$-factorization formula (\ref{a1}) with an infrared cut-off,
$k_T^2 > k_0^2$.  For $F_2^S$ we use the value of $F_2 (x, Q^2)$
obtained from the MRS(A$^\prime$) set of partons \cite{mrs} at $x
= 0.1$, and extrapolate below 0.1 assuming the normal $x^{-
0.08}$ \lq\lq soft" behaviour.

{}From Fig.\ 2 we see that the CCFM and DLLA predictions coincide
at large $x$, as indeed they should.  The two schemes start to
differ at small $x$ and Fig.\ 2 indicates the value of $x$ at
which the resummation effects become important.  It is evident
that once a background is added to the small $x$ behaviour
predicted by the CCFM equation then a good description of the
HERA data is obtained.  The prediction lies between the GRV and
MRS(A$^\prime$) values.  It should be noted that the CCFM
calculation is not a fit to the HERA data, but simply a solution
of the evolution equation incorporating angular ordering.  The
rise of the gluon, and hence of $F_2$, is generated by the
evolution equation and hence is within the domain of perturbative
QCD.  Of course the perturbative QCD prediction is not absolute.
The normalisation depends on the choice of $k_0^2$, which
delimits the infrared region, and also on the choice of the
driving term.  Also the normalisation depends on the choice of
the lower limit of integration in (\ref{a4}).  Here we take this
to be $\overline{Q}_0^2 = 1 \: {\rm GeV}^2$.  Recall that the
correction factor (\ref{a4}), and hence $\overline{Q}_0$, only
occurs because we solve a simplified form of the CCFM equation
appropriate to the small $x$ region.  In summary there is some
freedom in the normalisation of $F_2$, though the prediction of
the shape of the $x$ dependence is characteristic of the CCFM
equation.  It is encouraging that the physically reasonable
choice $k_0^2 = \overline{Q}_0^2 = 1 \: {\rm GeV}^2$ gives such a
satisfactory description of the HERA data.

For completeness we show in Figs.\ 3 and 4 respectively the
predictions for the longitudinal structure function $F_L (x,
Q^2)$ and for the charm component of $F_2$, which we denote by
$F_2^c (x, Q^2)$.  In each case we specify the background or
\lq\lq soft" contribution $F_i^S$ at $x = 0.1$ to be given by the
MRS(A$^\prime$) predictions, and extrapolate below 0.1 using the
$x^{- 0.08}$ \lq\lq soft" behaviour.  For the charm component
$F_2^c$ the CCFM predictions for $m_c = 1.4$ and 1.7 GeV are
shown; the argument of the running coupling is taken to be
$\kappa^2 + m_c^2$ (where $\kappa^2$ is shown in Fig.\ 1).  The
predictions of $F_L$ and $F_2^c$ obtained from GRV and
MRS(A$^\prime$) partons are also shown in Figs.\ 3 and 4.

Since the CCFM values of $F_2$ agree with the HERA data, we can
regard the charm component $F_2^c$ as an absolute prediction.
The charm component of the MRS(A$^\prime$) partons has been fixed
to be in agreement with the EMC measurements \cite{emc} of
$F_2^c$ which lie in the region $x \sim 0.1$.  Indeed we see
these data barely extend into the kinematic region shown in Fig.\
4.  It will be particularly informative to have measurements of
$F_2^c$ at HERA in the small $x$ regime where resummation effects
are expected to occur.

In summary we have shown that it is possible to obtain a good
description of the HERA measurements of $F_2$ from the solution
of a unified evolution equation based on the angular ordering of
the emitted gluons.  The gluon distribution $f (x, k_T^2,
\overline{Q}^2)$ was obtained by iteration starting from a
driving term of the form $3 (1 - x)^5 \exp (-k_T^2/k_a^2)$, and
the structure function $F_2$ was then determined via the
$k_T$-factorization formula $F_2 = f \otimes F_2^{\rm box}$.  The
steepness of the gluon, and of $F_2$, with decreasing $x$, is
generated by the evolution equation.  In this way we identified
the regime where the $\ln (1/x)$ resummations become important.

However, our treatment is only a first step.  There are several
reasons why it may overestimate the rise, particularly at low
$Q^2$.  First we have to find a realistic way to impose
energy-momentum conservation of the emitted gluons.  Second, we
have ignored gluon shadowing corrections.  These are expected to
be small in the HERA regime, as evidenced by the persistent rise
of the $F_2$ data with decreasing $x$ for $Q^2$ as low as $Q^2 =
2 \: {\rm GeV}^2$.  Last, but not least, the full next-to-leading
$\ln (1/x)$ contribution is unknown at present.  This is needed
to check the prescription for the running of $\alpha_S$ and to
specify the scale dependence.

Clearly the agreement of our CCFM predictions with the small $x$
measurements of $F_2$ do not imply angular ordering effects have
been firmly established.  GLAP and BFKL evolution can give an
equally good description.  There are two characteristic features
of the gluon distribution $f (x, k_T^2, \overline{Q}^2)$ obtained
{}from an evolution equation which includes a resummation of $\ln
(1/x)$ terms.  Namely a steep rise of $f$ with decreasing $x$
which is accompanied by a diffusion in $\ln k_T^2$.  $F_2$
measures only the rise.  A distinctive test will involve both
features.  For this we need to explore final state processes such
as deep inelastic events containing an identified energetic
forward jet.  Here we have focused on $F_2$ and obtained
predictions based on angular-ordered evolution which embodies
both BFKL and GLAP resummations.  Moreover, we have also
presented values for the charm component $F_2^c$ and the
longitudinal structure function $F_L$.  Valuable insight into the
properties of the gluon can be obtained from the measurement of
these structure functions at HERA.

\bigskip
\noindent {\large \bf Acknowledgements}

We thank Albert De Roeck, Mark Lancaster, Dick Roberts and
Andreas Vogt for useful information.  JK thanks the Department of
Physics and Grey College of the University of Durham for their
warm hospitality.  This work has been supported by Polish KBN
grant 2 P302 062 04 and the EU under contracts no.
CHRX-CT92-0004/CT93-0357.

\newpage

\noindent {\large \bf Figure Captions}
\begin{itemize}
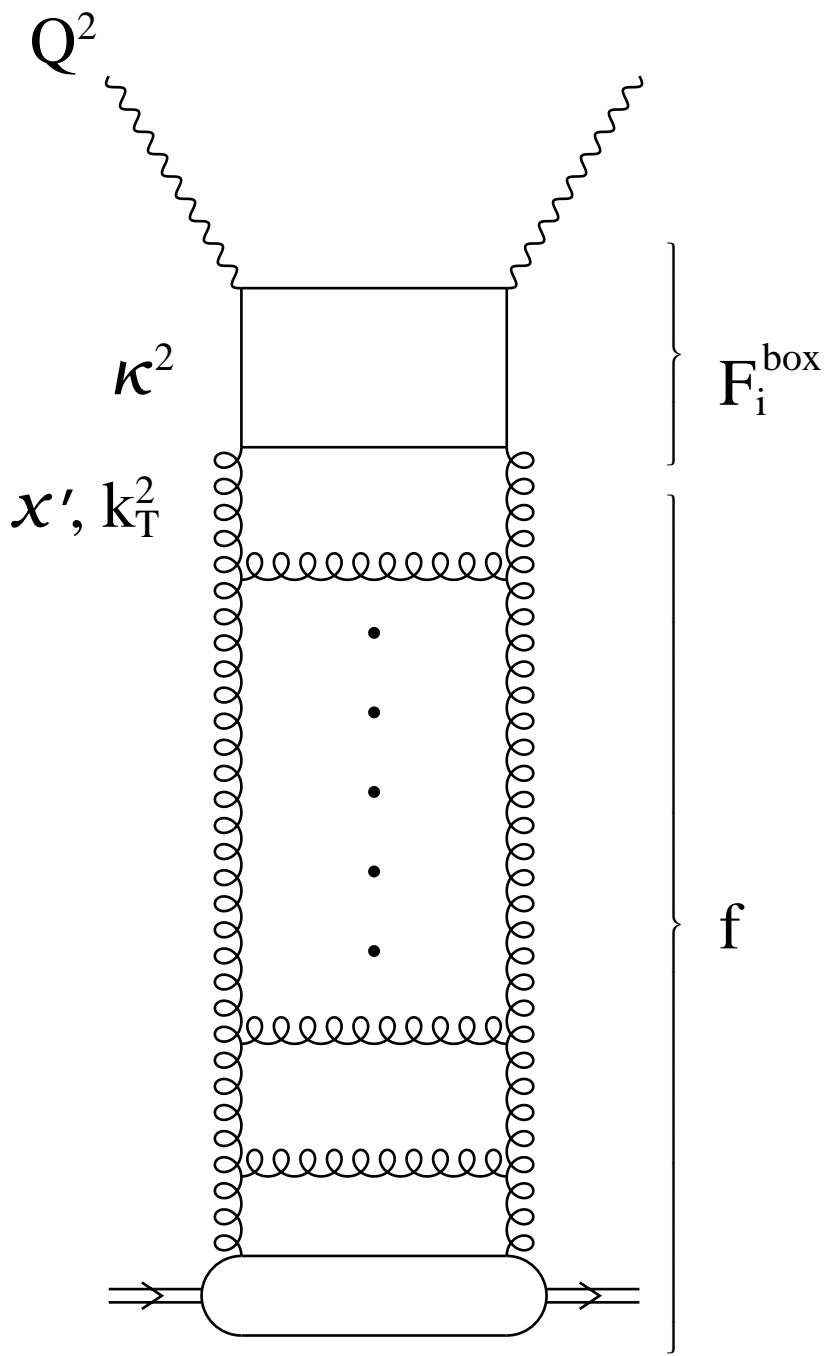
\item[Fig.\ 1] Pictorial representation of the
$k_T$-factorization formula, that is of the convolution $F_i = f
\otimes F_i^{\rm box}$ contained in eq.\ (\ref{a1}) with $i = 2,
L$.  $f (x^\prime, k_T^2)$ is the unintegrated gluon distribution
and $F_i^{\rm box}$ is the off-shell gluon structure function,
which at lowest order is determined by the quark box (and \lq\lq
crossed" box) contributions.  The integration variables,
$x^\prime$ and $k_T^2$, are respectively the longitudinal
fraction of the proton's momentum and the transverse momentum
carried by the gluon which dissociates into the $q\overline{q}$
pair.

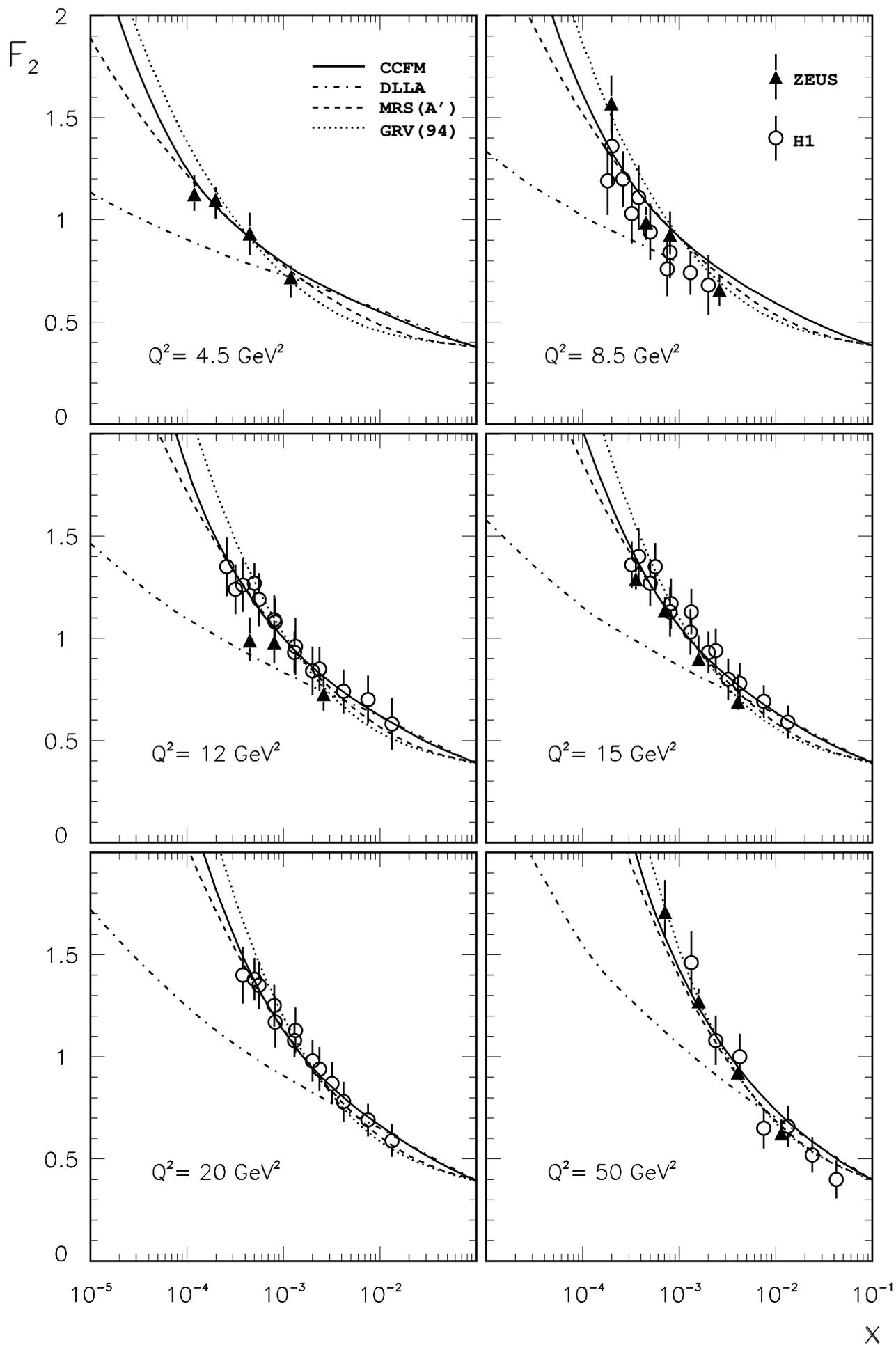
\item[Fig.\ 2] A comparison of the HERA measurements of $F_2$
\cite{h1,zeus} with the predictions
obtained from the
$k_T$-factorization formula (\ref{a1}) using for the unintegrated
gluon distribution $f$ the solutions of the CCFM equation
(continuous curve), and the DLL-approximation (dot-dashed curve)
of this equation.  We also show the values of $F_2$ obtained from
collinear factorization using the MRS(A$^\prime$) \cite{mrs} and
GRV \cite{grv} partons.

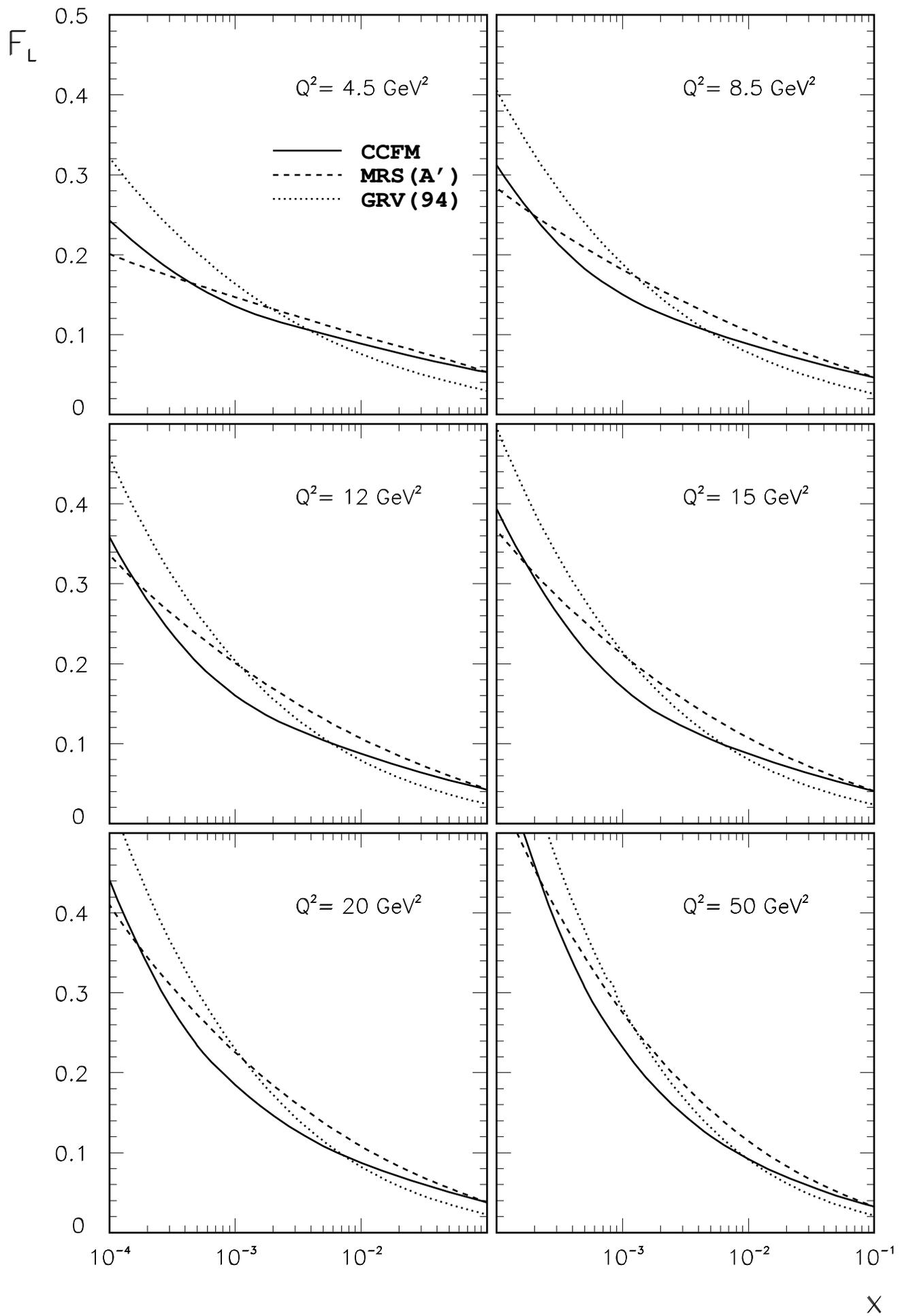
\item[Fig.\ 3] The continuous curve is the prediction for the
longitudinal structure function of the proton, $F_L (x, Q^2)$
obtained by solving the CCFM equation for the gluon and using the
$k_T$-factorization formula (\ref{a1}).  The values of $F_L$
obtained from GRV \cite{grv} and MRS(A$^\prime$) \cite{mrs}
partons are also shown.  The GRV prediction includes a charm
component only at leading order.

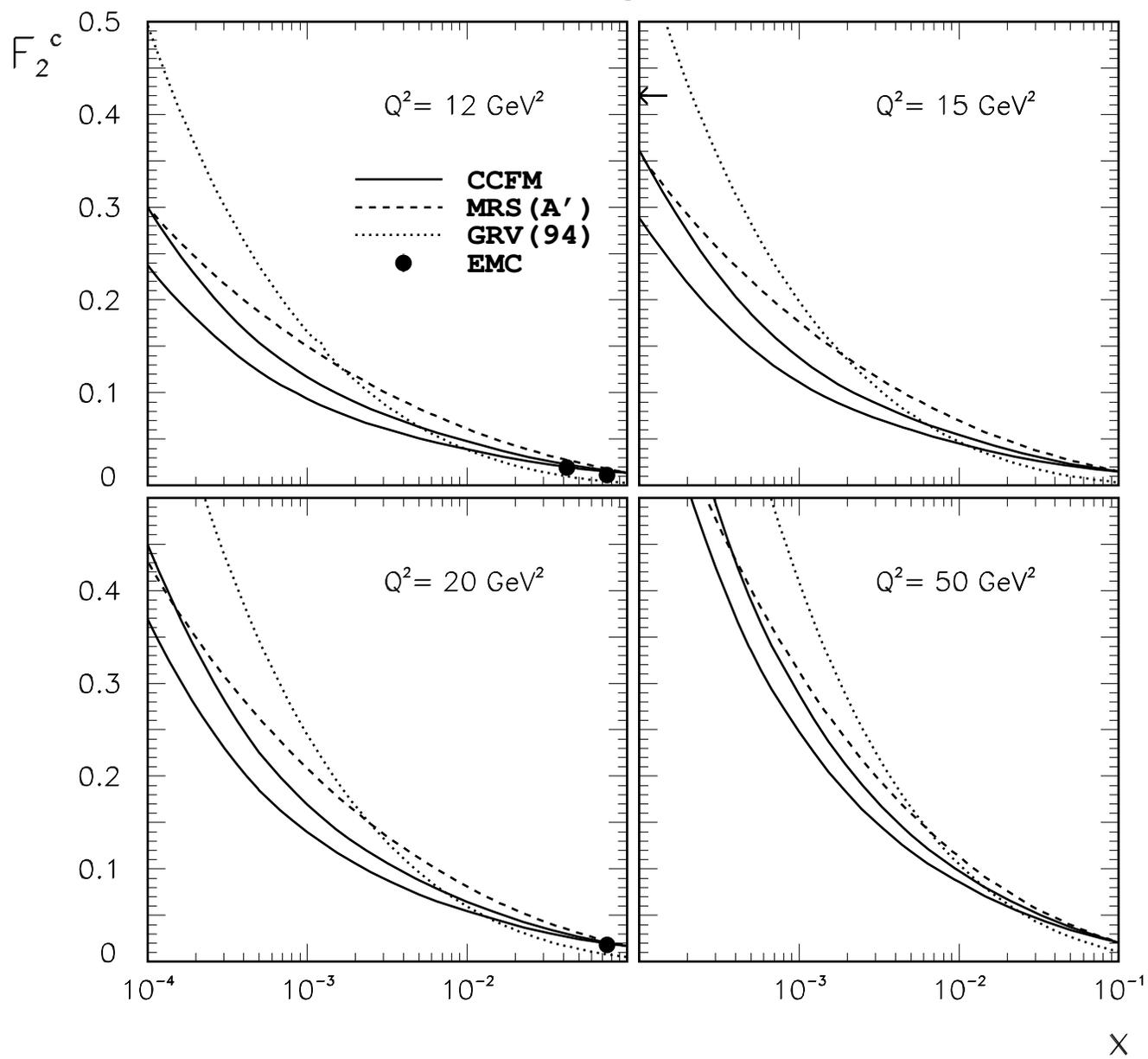
\item[Fig.\ 4] The continuous curves are the predictions for the
charm component $F_2^c$ of the proton structure function $F_2$
obtained by solving the CCFM equation for the gluon and using the
$k_T$-factorization formula with $m_c = 1.4$ (upper) and 1.7 GeV
(lower curve).  The values of $F_2^c$ obtained from GRV at
leading order \cite{grv} and from MRS(A$^\prime$) \cite{mrs}
partons are also shown.  The next-to-leading order GRV prediction
lies below the leading order result; for example at $Q^2 = 15 \:
{\rm GeV}^2$ and $x = 10^{- 4}$ it is shown by a small cross.
Also shown are EMC data \cite{emc} at adjacent $\langle Q^2
\rangle$ values, assuming that the $c \rightarrow \mu + X$
branching ratio is 8\%.

\end{itemize}
\end{document}